\documentclass{aa}
\usepackage{graphicx}
\usepackage{txfonts}
\usepackage{booktabs}
\usepackage{braket}
\usepackage[colorlinks=true, allcolors=blue]{hyperref}

\usepackage{graphicx}
\usepackage{pdfpages} 

\def\gsim{\mathrel{\spose{\lower 3pt\hbox{$\mathchar"218$}}
          \raise 2.0pt\hbox{$\mathchar"13E$}}}
\def\lsim{\mathrel{\spose{\lower 3pt\hbox{$\mathchar"218$}}
          \raise 2.0pt\hbox{$\mathchar"13C$}}}
          
\usepackage{verbatim,amscd, graphicx, amscd, mathrsfs}
\usepackage{amsmath}
\usepackage[english]{babel}
\usepackage[T1]{fontenc}
\usepackage{color}
\usepackage{listings}
\usepackage{lmodern}
\usepackage[colorlinks=true, allcolors=blue]{hyperref}
\usepackage{caption}
\usepackage{geometry}
\usepackage{graphicx}
\usepackage{multicol}
\usepackage{multirow}
\usepackage{float}
\usepackage{braket}
\usepackage{ulem}
\usepackage[utf8]{inputenc}

\begin{document}

\title{Ending the prompt phase \\in photospheric models of gamma-ray bursts}

\author{Filip Alamaa$^{1,2}$, Fr\'ed\'eric Daigne$^{1,3}$, and Robert Mochkovitch$^{1}$}

\authorrunning{F. Alamaa et al.}
\titlerunning{ESD in photospheric GRBs}

\institute{\textsuperscript{1}Sorbonne Université, CNRS, UMR 7095, Institut d'Astrophysique de Paris (IAP), 98 bis boulevard Arago, 75014 Paris, France\\
\textsuperscript{2}Department of Physics, KTH Royal Institute of Technology, and The Oskar Klein Centre, SE-10691 Stockholm, Sweden\\
\textsuperscript{3}Institut Universitaire de France, Ministère de l'Enseignement Supérieur et de la Recherche, 1 rue Descartes, 75231 Paris Cedex F-05, France}

\abstract{The early steep decay, a rapid decrease in X-ray flux as a function of time following the prompt emission, is a robust feature seen in almost all gamma-ray bursts with early enough X-ray observations. This peculiar phenomenon has often been explained as emission from high latitudes of the last flashing shell. However, in photospheric models of gamma-ray bursts, the timescale of high-latitude emission is generally short compared to the duration of the steep decay phase, and hence an alternative explanation is needed. In this paper, we show that the early steep decay can directly result from the final activity of the dying central engine.
We find that the corresponding photospheric emission can reproduce both the 
temporal and spectral evolution observed. This requires a late-time behaviour that should be common to all GRB central engines, and we estimate the necessary evolution of the kinetic power and the Lorentz factor. 
If this interpretation is correct, observation of the early steep decay 
can grant us insights into the last stages of central activity, and provide new constraints on the late evolution of the Lorentz factor and photospheric radius.
}


\maketitle
\section{Introduction}
The early steep decay (ESD) is a common feature in gamma-ray bursts (GRBs) light curves. It is observed in X-rays during the transition between the GRB prompt phase and the subsequent GRB afterglow phase. 
At the end of the violent prompt phase, the observed luminosity in X-rays drops rapidly several orders of magnitude with a temporal index of $-3$ to $-5$. Typical durations for the ESD are $\sim 10^2-10^3~$s  \citep{Nousek2006}. This behaviour seems to be a robust feature in GRBs: if an X-ray telescope (such as the XRT on board \textit{Swift}) manages to observe a GRB early enough, this peculiarity is almost always seen.

A natural explanation is that the ESD is the consequence of high-latitude emission (HLE) from the last flashing shell in the optically thin regime \citep{Fenimore1996, KumarPanaitescu2000, GenetGranot2009}. In this case, the bolometric luminosity, $L_{\rm bol}$, and the peak energy of the spectrum, $E_{\rm p}$, are expected to decrease with time approximately as $L_{\rm bol}\propto t^{-3}$ and $E_{\rm p} \propto t^{-1}$, respectively.
This theoretical expectation was recently observationally corroborated by \citet{Tak2023} \citep[see also][]{Uhm2022arXiv}. By performing time-resolved fits of broad GRB prompt pulses, the authors found a parameter evolution consistent with that expected from HLE in a majority of the pulses examined. Given that HLE emission is observed in the decay phases of prompt pulses, as found by \citet{Tak2023}, one could argue that it is plausibly the origin for the ESD as well, since there is often a smooth transition between the last prompt pulse and the ESD. 

By contrast, \citet{Ronchini2021} performed time-resolved analysis during the ESD using data from XRT and found that the spectral evolution does not match that predicted for HLE. During the observations, the peak energy seems to cross the whole XRT band, indicating a stronger spectral evolution as $E_{\rm p}\propto t^{-2}$ to $t^{-2.5}$. By fitting a power-law spectrum to the XRT data, they found a relation between the fitted spectral index, $\alpha(t)$, and the ratio of the maximum flux at the onset of the decay to the flux
at time $t$, $F_{\rm max}/F(t)$. \citet{Ronchini2021} referred to this correlation as the ``$\alpha-F$ relation'', which they interpreted as due to adiabatic cooling of the emitting particles in the context of a proton synchrotron origin of the GRB emission. 

In photospheric models of GRBs, the prompt radiation is emitted when the initially opaque jet transitions to the optically thin regime \citep{Paczynski1986, Goodman1986}. 
This transition usually occurs once the acceleration of the outflow is complete, at a
characteristic radius, 
$R_\mathrm{ph}$,  given by \citep[e.g.,][]{Hascoet2013}
%
\begin{equation}\label{eq:Rph}
       R_{\rm ph}=\frac{f \kappa_{\rm T} {\dot E}}{
    8\pi c^3 (1+\sigma) \Gamma^3}
    =5.9\times 10^{12}\, \frac{f \dot{E}_{52}}{(1+\sigma)\Gamma_2^3}\ \ {\rm cm}.
\end{equation}
Here, $\dot{E}$ is the isotropic equivalent total injected power in the outflow,
$\Gamma$ is the bulk Lorentz factor,
$\sigma$ is the magnetization at large distances,
$f$ is the number of leptons per baryon, and $\kappa_{\rm T} = 0.4~$cm$^2~$g$^{-1}$ is the Thomson opacity. 
In the following, we assume a negligible magnetization at large distances ($\sigma\ll 1$).
The total power, $\dot{E}$, is equal to the kinetic power, $\dot{E}_\mathrm{K}$, in this case.
The corresponding geometrical timescale is given by
\begin{equation}\label{eq:tau_geo}
    \tau_{\rm geo}=\frac{R_{\rm ph}}{2c \Gamma^2}=\frac{f \kappa_{\rm T} {\dot E}_{\rm K}}
    {16\pi c^4 \Gamma^5}=9.8\times 10^{-3}\,  \frac{f{\dot E}_{{\rm K},52}}{\Gamma_2^5}\ \ {\rm s},
\end{equation}
which is very short, except if $\Gamma$ is small, or ${\dot E}_{\rm K}$ or $f$ is huge \citep[see e.g.,][]{Dereli2022, SamuelssonRyde2023}. 

As discussed by \citet{Hascoet2012}, 
a short geometrical timescale
argues against a HLE interpretation for the ESD in photospheric models 
(or the decay observed at the end of prompt pulses), since the total duration of the ESD should be of the order $\sim \tau_{\rm geo}$ \citep[however, see][for a longer-lasting ESD originating from photons diffusing in the surrounding cocoon]{Peer2006ESD}. Furthermore, $\tau_{\rm geo}\propto \Gamma^{-5}$, which is very sensitive to $\Gamma$ while the ESD appears to be a robust feature,
nearly always present at the start of the early afterglow of GRBs with a generic temporal decay index of $-3$ to $-5$ \citep{Nousek2006}. 

In light of the above argumentation, we 
investigate in this paper an alternative origin for the ESD in photospheric models, namely that it results from the intrinsic evolution of the dying central engine \citep[as suggested by][]{Hascoet2012}. Depending on the last stages in the life of the central source, the photospheric emission may indeed generate an ESD mimicking that expected from HLE or a stronger spectral evolution as the one found by \citet{Ronchini2021}. Assuming a power-law decline in the emitted power and Lorentz factor of the flow, we estimate the observed light curve and spectrum under different scenarios and compare the results to observations.

The paper is structured as follows. In Sect. \ref{Sec:methodology}, we introduce our model for the last stages of central activity. To predict the observed signal, we study two benchmark scenarios: a non-dissipative model in Sect. \ref{sec:non_diss} and a dissipative model in Sect. \ref{sec:constant_diss}. In Sect. \ref{Sec:results}, we present our results and discuss them, with a specific emphasis on parameter dependence and underlying assumptions, in Sect. \ref{Sec:discussion}. We conclude in Sect. \ref{sec:conclusion}. We employ the notation $Q_x = Q/10^x$ throughout the text.

\section{The ESD in photospheric models}\label{Sec:methodology}
We model the last stages of the central engine with power-law decays for the injected power and Lorentz factor of the relativistic wind as
\begin{equation}\label{eq:time_evolution}
    {\dot E}_{\rm K}={\dot E}_{\rm b}\,\left(\frac{t}{t_{\rm b}}\right)^{-\lambda} ,\ \ \ \Gamma=\Gamma_{\rm b}\,\left(\frac{t}{t_{\rm b}}\right)^{-\gamma}, \qquad t > t_{\rm b}
\end{equation}
where $t$ is the time in the source rest frame, $t_{\rm b}$ is the 
break time when the ESD starts, and ${\dot E}_{\rm b}$ and $\Gamma_{\rm b}$ are, respectively, the kinetic power and Lorentz factor at $t_{\rm b}$. The choice of power-law decays for ${\dot E}_{\rm K}$ and $\Gamma$ is discussed in Sect. \ref{sec:progenitor}.

If the emitted bolometric luminosity follows the evolution of ${\dot E}_{\rm K}$ given in the equation above, i.e., if the radiative efficiency does not vary too much along the ESD, then one expects $\lambda \sim 3$ to account for ESD light curves. If, in addition, $f$ does not vary too much during the ESD either, one finds from Eq. \eqref{eq:Rph} that the photospheric radius evolves as $R_\mathrm{ph} \propto t^{3(\gamma-1)}$. Thus, the photosphere can either shrink or inflate with time depending on the value of ($\gamma$-1).


\subsection{Non-dissipative model}\label{sec:non_diss}
In a non-dissipative photospheric model, the radiation below the photosphere is at all times kept in thermodynamic equilibrium with the plasma. Under this assumption, the temperature of the observed radiation and the observed isotropic equivalent 
bolometric luminosity at the photosphere are given by \citep[e.g.,][]{Piran1999, DaigneMochkovitch2002, Peer2007}
\begin{equation}\label{eq:T_obs}
    T_{\rm obs} = T_0 \left(\frac{R_{\rm ph}}{R_s}\right)^{-2/3}, \quad L_{\rm bol}
= {\dot E}_{\rm th}
    \left(\frac{R_{\rm ph}}{R_s}\right)^{-2/3}.
\end{equation}
Here, $T_0$, ${\dot E}_{\rm th}$, and $R_s=\Gamma R_0$
are, respectively, the initial temperature, the injected power in thermal form at the base of the jet, and the saturation radius, and
$R_0$ is the distance from the central engine at the base of the jet. 
The initial temperature is given by
\begin{equation}\label{eq:T_0}
  kT_0 \approx k\left(\frac{{\dot E}_{\rm th}}{4\pi R_0^2 a c}\right)^{1/4}= 1.2\,{\dot E}_{{\rm th},52}^{1/4}\,R_{0,7}^{-1/2}\ \ {\rm MeV}.
\end{equation}
We assume that thermal energy has been efficiently converted into kinetic energy below the photosphere, such that we have ${\dot E}_{\rm th} \approx {\dot E}_{\rm K}$.

The efficiency of the non-dissipative model can be evaluated from Eqs. \eqref{eq:Rph} and (\ref{eq:T_obs})
\begin{equation}
\epsilon_\gamma=\frac{L_{\rm bol}}
{{\dot E}_{\rm th}}=\left(\frac{R_{\rm ph}}{R_s}\right)^{-2/3}=3\times 10^{-3}\,\left(\frac{f{\dot E}_{{\rm K},52}}{R_{0,7}\Gamma_2^{4}}\right)^{-2/3},
\end{equation}
which is quite small for standard values of the parameters. Similarly, the observed temperature, equal to $T_0\epsilon_\gamma$, is expected in the keV range. 

Due to HLE and contributions from different optical depths to the released emission, the observed spectrum consists of a superposition of black bodies at different temperatures. This leads to a modified, broadened spectrum compared to a Planck function in the observer frame. This spectrum can be approximated by a cutoff power-law function with a spectral index $\alpha = 0.4$ and a peak energy of $E_{\rm p}^{\rm th} = 3.9kT_{\rm obs}$ \citep{Beloborodov2010}. Assuming $\lambda = 3$ and that $R_0$ and $f$ do not vary during the ESD, a HLE-like decline of the peak energy with time as $t^{-1}$ requires $\gamma = 27/32 \sim 0.8$.


\subsection{Dissipative model}\label{sec:constant_diss}
Energy injection below the photosphere provides a way to increase the efficiency, raise the peak energy of the emitted spectrum, and transform its shape via Comptonization. In this section, we consider an unspecified dissipation mechanism that continuously injects energy into the photon distribution. Such a mechanism could be for instance magnetic dissipation \citep{Drenkhahn2002, Giannios2008}, long-lasting dissipation via turbulence or multiple shocks \citep{ReesMeszaros2005, Zrake2019}, or collisional heating between neutrons and protons \citep{Beloborodov2010}. We choose not to detail the processes involved and rather adopt some very simplified assumptions to estimate the spectral evolution and the outflow parameters during the ESD. 

The bolometric luminosity is obtained by fixing the efficiency such that 
\begin{equation}\label{eq:L_gamma_diss}
L_{\rm bol} =\epsilon_\gamma {\dot E}_{\rm K},
\end{equation}
with $\epsilon_\gamma = 0.1\ -\ 0.5$. 
The efficiency can be large if dissipation takes place close to the photospheric radius \citep{Beloborodov2010, Gottlieb2019, SamuelssonRyde2023}. Although the efficiency could very well vary during the ESD, here we keep it constant for simplicity. This is somewhat in line with the numerical simulation presented in \citet{Gottlieb2019}, where $\epsilon_{\gamma}$ was found to fluctuate around a central value of $\epsilon_{\gamma} \sim 0.5$. In this paper, we model the global behaviour and neglect possible small scale variations in ${\dot E}_{\rm K}$, $\Gamma$, and/or $\epsilon_{\gamma}$, which we argue can be at least partially suppressed (see Sect. \ref{sec:blending} below). If there exists a global time-dependence for $\epsilon_{\gamma}$, the results in presented in Sect. \ref{Sec:results} can still be obtained by adjusting $\lambda$ and $\gamma$ accordingly. 

If the bolometric radiation came from a blackbody,\footnote{Note that the effective temperature $T_\mathrm{eff}$ in the dissipative case differs from the value found in the non-dissipative case due to a different radiated power $L_\mathrm{bol}$.}
the peak energy would be related to the 
effective temperature, $T_{\rm eff}$, as \citep{Beloborodov2013} 
\begin{equation}\label{eq:Ep}
\begin{split}
    E_{\rm p}^{\rm th} & \approx 4\Gamma kT_{\rm eff} = 44 \ L_{{\rm bol},52}^{1/4} \left(\frac{\Gamma_2}{R_{{\rm ph},13}}\right)^{1/2} \ {\rm keV}\\
    &= 57 \ L_{{\rm bol},52}^{-1/4}\,\Gamma_2^2\,\left(\frac{\epsilon_\gamma}{f}\right)^{1/2}
\ {\rm keV},
\end{split}
\end{equation}
where the third equality employs Eq. \eqref{eq:L_gamma_diss}. 
We assume that the Comptonized spectrum satisfies $E_{\rm p} \sim {\rm a\,few}\ E_{\rm p}^{\rm th}$, as argued in \citet{Beloborodov2013}. If, in addition, we adopt the simplifying assumptions that $\epsilon_\gamma$, $f$, and the ratio $E_{\rm p}/E_{\rm p}^{\rm th}$ do not vary too much during the ESD, we find from Eq. \eqref{eq:Ep} that the peak energy follows a power-law 
\begin{equation}
E_{\rm p}\propto {\dot E}_{\rm K}^{-1/4}\,\Gamma^2
\propto t^{\lambda/4-2\gamma}\ .
\end{equation}
Adopting $\lambda=3$, a HLE-like behaviour, i.e., $E_{\rm p}\propto t^{-1}$, is obtained for $\gamma=7/8$, while the strongest spectral evolution found by \citet{Ronchini2021}
requires $\gamma=1.4$ to 1.6.

Lastly, we assume that the dissipation transforms the observed spectrum to be similar to that of typical GRB observations. Specifically, we consider the observed spectrum to be a Band-function \citep{Band1993}, with a soft, low-energy power-law index $\alpha = -1$ and a high-energy power-law index $\beta= -2.3$. We discuss the influence of this choice on the results in Sect. \ref{sec:parameter_dependence}.

\section{Results}\label{Sec:results}
\begin{figure*}
    \centering
    \includegraphics[width=0.45\linewidth]{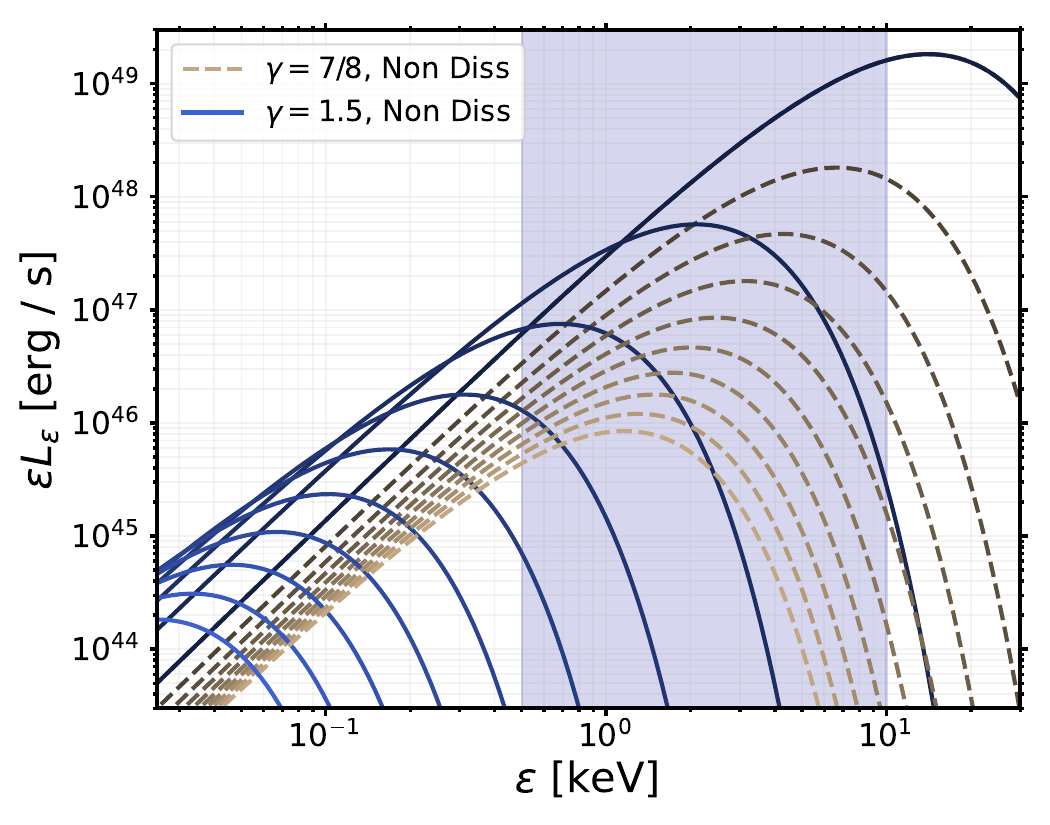}
    \includegraphics[width=0.45\linewidth]{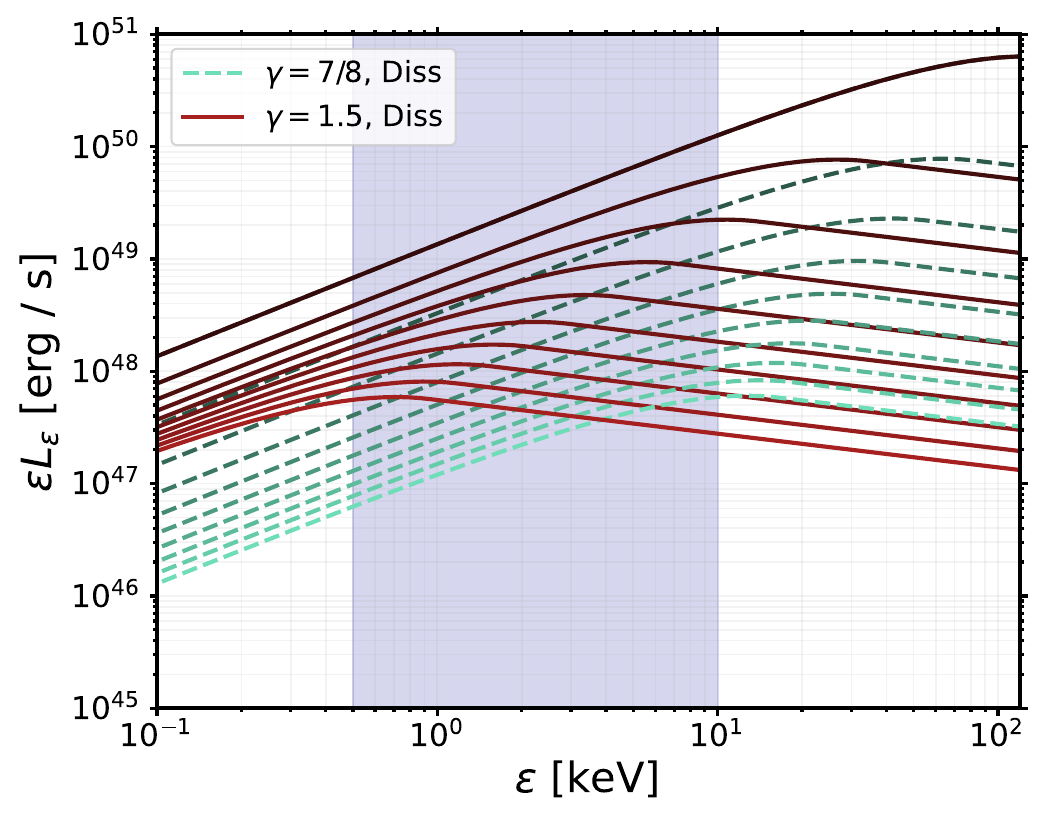}\\
\caption{Snapshot spectra showing the evolution of the spectral luminosity during the ESD in the non-dissipative scenario (left) and the dissipative scenario (right). Dashed lines have $\gamma = 7/8$, while solid lines have $\gamma = 1.5$, as also indicated in the figure. The spectra are plotted at even intervals in time from the onset of the ESD at $t/t_b = 1$ (dark color) until $t/t_b = 10$ (bright color). Note that the two spectra for $\gamma = 7/8$ and $\gamma = 1.5$ are overlapping at $t/t_b = 1$. The purple shading indicates the energy sensitivity of XRT. Parameters used are given in Table \ref{tab:values}.}
\label{fig:spectra}
\end{figure*}

\begin{figure}
\centering
\includegraphics[width=0.9\linewidth]{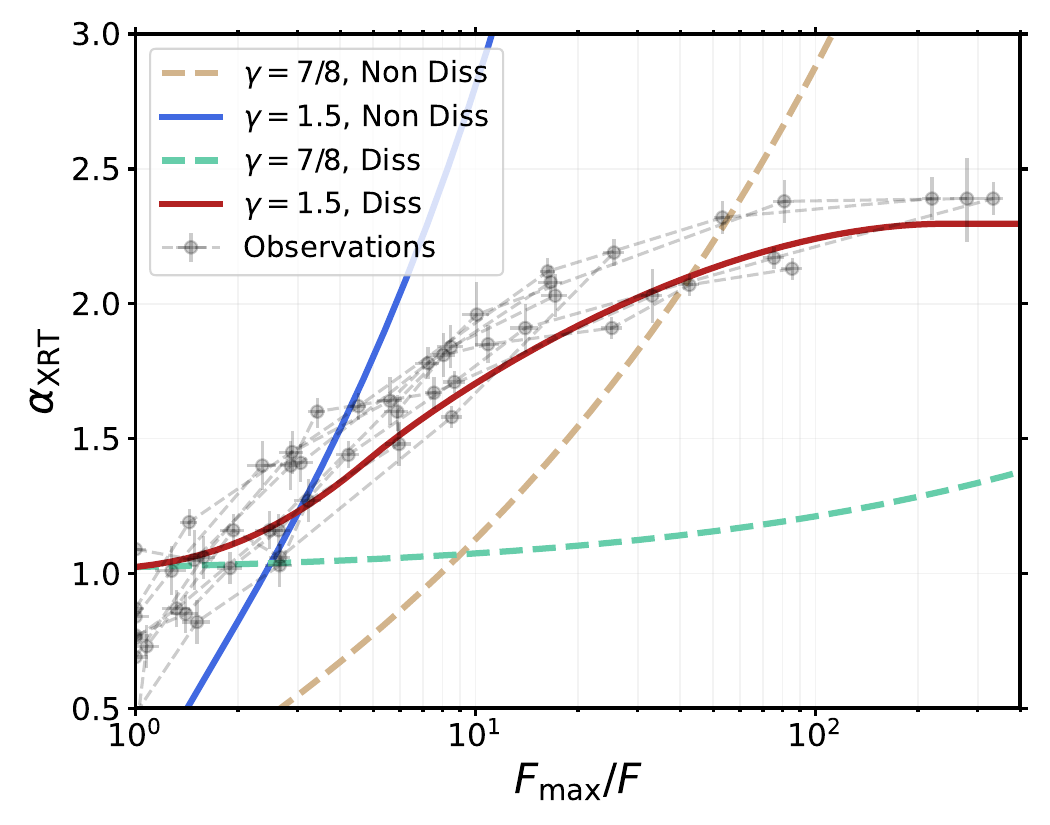}
\caption{Photon index as a function of $F_{\rm max}/F$ along the ESD branch. 
The grey points are the best-fit values from time-resolved analyses of 8 GRBs, as obtained in \citet{Ronchini2021}. Parameter values used are given in Table \ref{tab:values}.}
\label{fig:alpha_F}
\end{figure}

\subsection{Spectral evolution during the ESD}
In Fig. \ref{fig:spectra}, the evolution of the spectral shape during the ESD is shown, in the non-dissipative (left panel) and in the dissipative scenario (right panel). Both scenarios are shown for $\gamma = 7/8$ (dashed lines, intended to mimic a HLE-like evolution), and $\gamma = 1.5$ (solid lines).
As expected, when $\gamma = 1.5$, the decrease of $E_{\rm peak}$ is much faster compared to when $\gamma = 7/8$.

In Fig. \ref{fig:alpha_F}, we show the obtained $\alpha-F$ relation. The maximum flux, $F_{\rm max}$, is calculated as the integrated spectral flux within the XRT energy band ($0.5-10~$keV) at the onset of the ESD (first spectrum in Fig. \ref{fig:spectra}), and $F(t)$ is the flux in the XRT-band at time $t$. The spectral index is estimated as in \citet{Ronchini2021}:
\begin{equation}\label{eq:alpha}
    \alpha_{\rm XRT} = -\frac{\log[N_{\nu = 10\, {\rm keV}/h}(t) \, /\, N_{\nu = 0.5\, {\rm keV}/h}(t)]}{\log(10 \, {\rm keV}/0.5 \, {\rm keV})}.
\end{equation}
where $N_{\nu = E/h}$ is the spectral number density, evaluated at energy $E$.

The line-coding in Fig. \ref{fig:alpha_F} is the same as for the corresponding scenarios in Fig. \ref{fig:spectra}. The grey points show the best-fit values obtained in \citet{Ronchini2021} in their time-resolved spectral fits during the ESD of 8 GRBs. It is clear from the figure that unless there is some dissipation occurring below the photosphere, the spectrum is initially too hard and at late times too soft to account for the observations. 
The \citet{Ronchini2021} results are well 
reproduced
in the dissipative case with $\gamma = 1.5$ (full red line). The peak energy then crosses the entire XRT band, starting at $E_{\rm p} \gtrsim 100~$keV and reaching 0.5~keV, the lower limit of the XRT energy band considered here, at $F_{\rm max}/F \sim 150$ ($t/t_b \gtrsim 10$, see Fig. \ref{fig:spectra}). Conversely, $\gamma=7/8$, which mimics the spectral evolution during HLE, fails (by far) to reproduce the \citet{Ronchini2021} results (dashed green line).
It should however be noted that the \citet{Ronchini2021} results rely on the delicate correction for absorption, which becomes quite large 
below 1 keV (see discussion in Sect. \ref{sec:caution} below).
\begin{table}[t]
    \centering
    \caption{Parameter values used in Figures \ref{fig:spectra}, \ref{fig:alpha_F}, and \ref{fig:parameter_evolutions}, unless otherwise stated}
    \begin{tabular}{lll}
    \hline
         Parameter & Non-dissipative & Dissipative \\
    \hline
         ${\dot E}_{\rm b}$ & $10^{52}~$erg~s$^{-1}$ & $10^{52}~$erg~s$^{-1}$ \\
         $\Gamma_{\rm b}$ & 100 & 100 \\
         $\lambda$ & 3 & 3 \\
         $t_b$ & 20~s & 20~s \\
         $f$ & 1 & 1 \\
         $R_0$ & $10^7~$cm & - \\
         $\epsilon_\gamma$ & - & 0.3 \\
         $E_{\rm p}/E_{\rm p}^{\rm th}$ & - & 3 \\
         $\alpha$ & 0.4 & $-1$ \\
         $\beta$ & - & $-2.3$ \\
         $z$ & 1 & 1 \\
    \hline
    \end{tabular}
    \label{tab:values}
\end{table}

\subsection{Outflow parameters during the ESD}
\begin{figure*}
    \centering
    \includegraphics[width=0.4\linewidth]{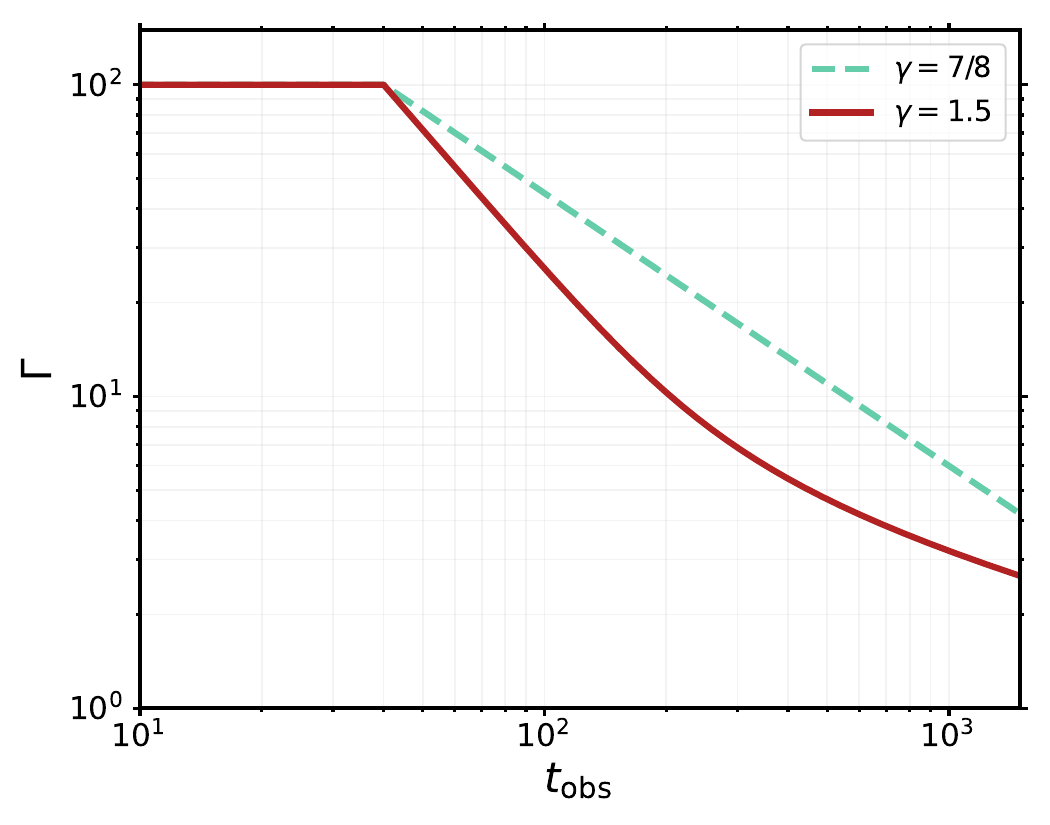}
    \includegraphics[width=0.4\linewidth]{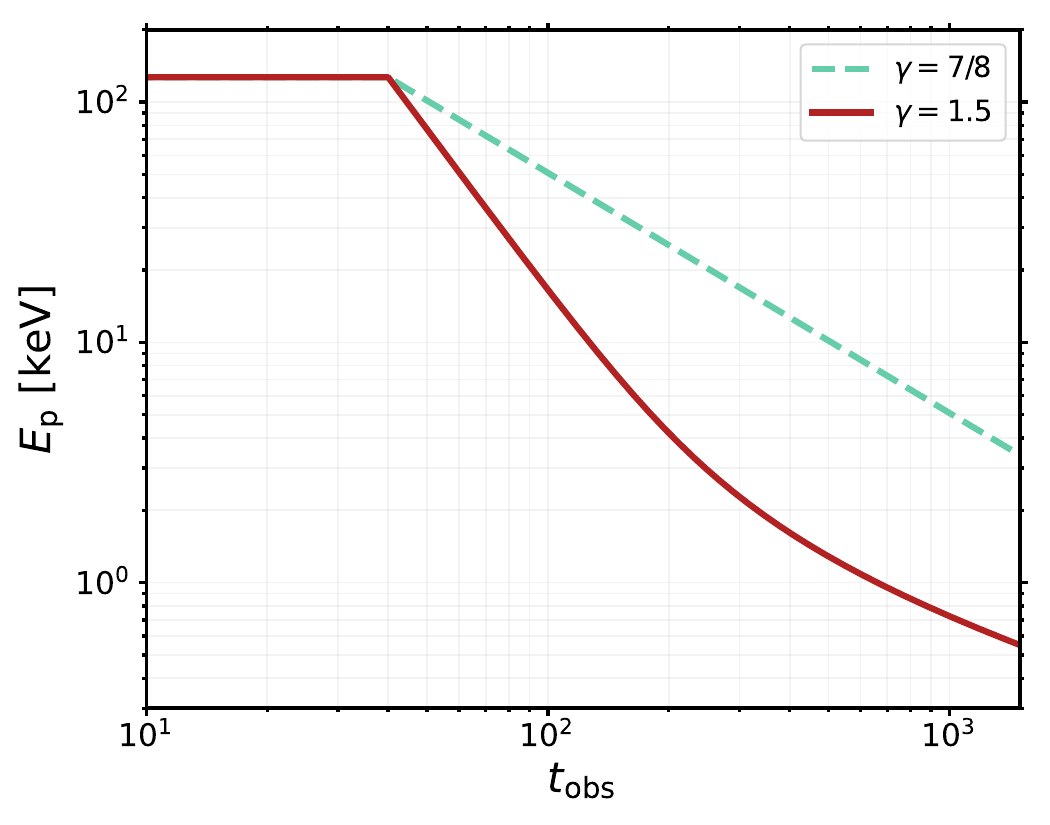}\\
    \includegraphics[width=0.4\linewidth]{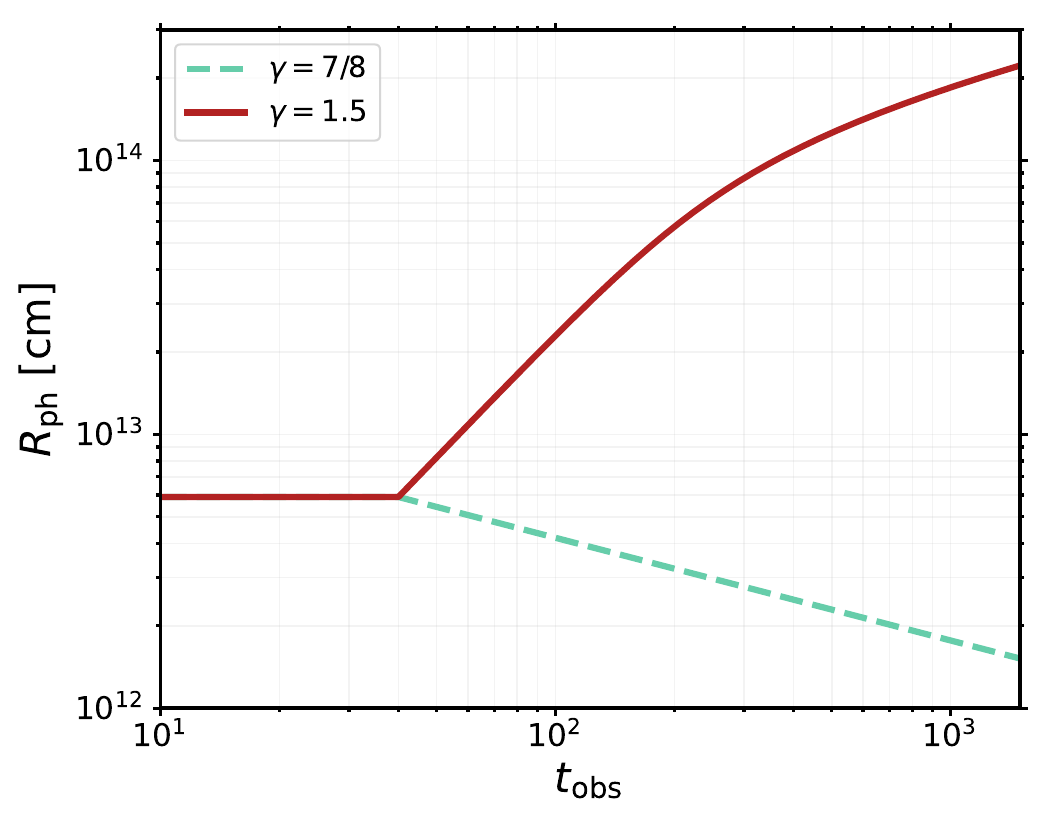}
    \includegraphics[width=0.4\linewidth]{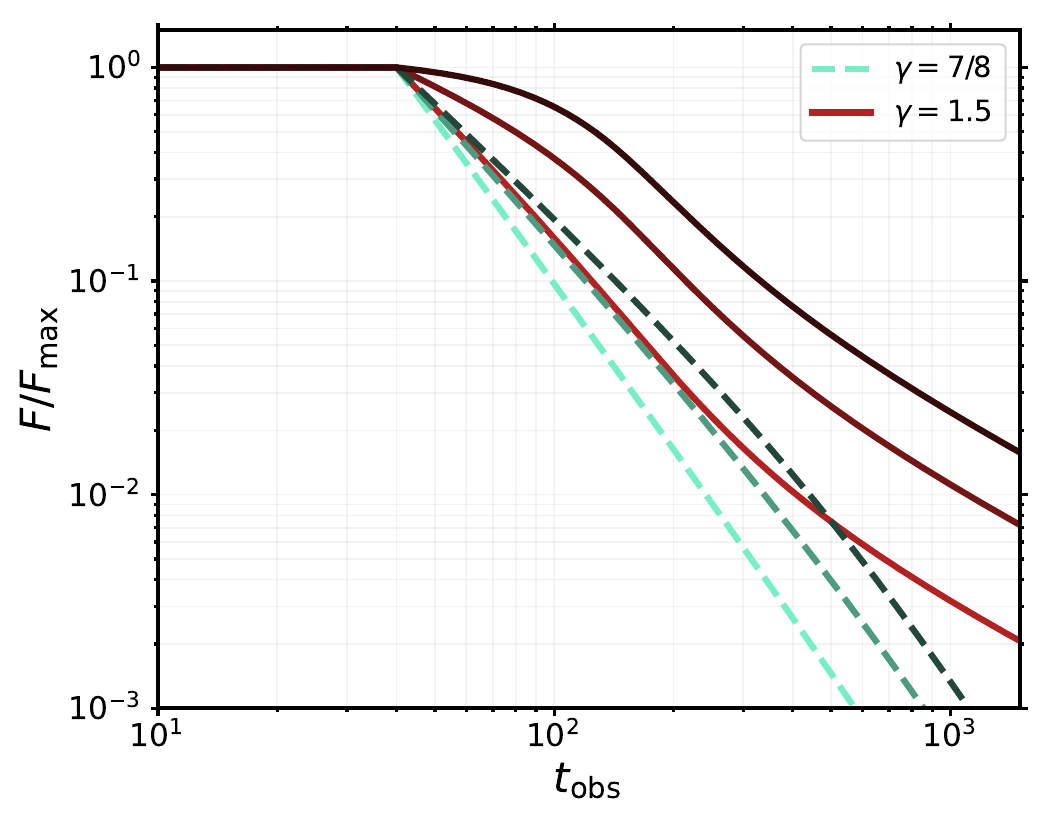}
\caption{Evolution of the Lorentz factor, peak energy, photospheric radius, and normalized XRT-flux during the ESD for the dissipative model. The XRT-flux in the bottom-right panel is shown for the low-energy index $\alpha$ equal to $-2/3$ (dark color), $-1$ (intermediate color), and $-1.5$ (bright color).  Parameter values are given in Table \ref{tab:values}.}
\label{fig:parameter_evolutions}
\end{figure*}

The evolution of the Lorentz factor $\Gamma$, peak energy $E_{\rm p}$, and photospheric radius $R_{\rm ph}$ in the dissipative scenario are shown in Fig. \ref{fig:parameter_evolutions}. The quantities are plotted against observer time
\begin{equation}
    t_{\rm obs}=(1+z) [t+\tau_{\rm geo}],
\end{equation}
where $\tau_{\rm geo}$ appears to account for the propagation time of the plasma to the photosphere: $t_{\rm prop} = R_{\rm ph}/2c\Gamma^2 = \tau_{\rm geo}$, assuming emission on the line-of-sight. For $\gamma=7/8$, $\tau_{\rm geo}$ 
is always negligible and $\Gamma$, $E_{\rm p}$ and 
$R_{\rm ph}$ just follow power-laws of slopes $-7/8$, $-1$, and $-3/8$, respectively. For $\gamma=1.5$, the quantities start decaying as power laws of slopes $-1.5$, $-2.25$, and $+1.5$, from which they deviate once $\tau_{\rm geo}$ becomes comparable to $t$. 

One can estimate when this transition occurs by setting $t = \tau_{\rm geo}$. This gives
\begin{equation}
    t = t_{\rm HLE} \equiv \left( t_{\rm b}^{5\gamma - \lambda} \, \frac{100 \, \Gamma_{{\rm b},2}^5}{f \, {\dot E}_{{\rm b},52}} \right)^{1/(5\gamma - \lambda - 1)},
\end{equation}
with the above equation being valid only when $5\gamma - \lambda > 1$. Adopting $\lambda = 3$, $\gamma = 1.5$, $f= 1$, and $t_b = 20~$s, we obtain $t_{\rm HLE} = 175~$s, which, with $z=1$, gives $t_{\rm obs} = 700~$s. This is the time when HLE starts to dominate. From the figure, it is clear that the deviations become noticeable earlier than this once $t_{\rm obs}\sim 200~$s, which corresponds to $\tau_{\rm geo}/t \sim 0.1$.

When $t>t_{\rm HLE}$, HLE emission can become important and our results should be evaluated with caution since we only consider emission on the line-of-sight. HLE from the optically thick fireball results in the bolometric flux decreasing as $t^{-2}$ with a slowly evolving peak-energy \citep{PeerRyde2011}. Note that when $t>t_{\rm HLE}$, the observer would still see HLE from the last emitted regions even if the central engine activity ceased. 

Finally, the ESD light curve (in the XRT spectral band 0.5 - 10 keV) is shown in the bottom right panel in Fig. \ref{fig:parameter_evolutions}. We represent the light curves for three values of the low-energy spectral index: $\alpha=-2/3$, $-1$, and $-1.5$. When $\gamma = 7/8$, the flux immediately starts to decrease rapidly after $t_{\rm b}$ due to the drop in $L_{\rm bol}$. For $\gamma = 1.5$ however, the decay is more gradual. This is because the drop in luminosity is partially prevented by the peak energy moving into the XRT-band, as also evident from Fig. \ref{fig:spectra}. 
Both these types of behaviours have been observed. For instance, GRB 090618 and GRB 230420A show very steep decays right after the prompt emission, while GRB 081221 and GRB 210305A show more smooth transitions to the ESD phase, similar to the solid black line in Fig. \ref{fig:parameter_evolutions}.\footnote{XRT light curves are available at \url{https://www.swift.ac.uk/xrt_curves/}.}




\section{Discussion}\label{Sec:discussion}
\subsection{Parameter dependence}\label{sec:parameter_dependence}
There are two observables that we can compare our predictions against during the ESD, the typical XRT-light curve and the spectral evolution. The XRT-light curve commonly drops $\sim 2$ order of magnitude in $\sim 100$~s, seemingly as a power law. As evident from the bottom right panel in Fig. \ref{fig:parameter_evolutions}, the flux decrease predicted by our model depends on the value of $\gamma$ as well as on the spectral shape. However, it depends most strongly on the choice of $\lambda$; a larger $\lambda$ leads to a more rapid decrease. The drop in flux can be circumvented to an extent by a different choice of $\gamma$ and/or a different spectral shape, but this requires fine-tuning once $\lambda > 5$. The same is true for $\lambda < 3$. Thus, we deduce that the isotropic equivalent of the kinetic power must drop as a power law in time with index ~$\sim 3-$5 to account for the observed XRT-light curve. If this scenario is correct,
this provides an important constraint on the behaviour of the central engine at the end of the relativistic jet launching (see discussion in Sect. \ref{sec:progenitor}).

The results regarding the spectral evolution presented in Fig. \ref{fig:alpha_F} are most sensitive to the assumed shape of the emitted spectrum. The general trend 
in the $\alpha$--$F$ plane (Fig.~\ref{fig:alpha_F})
is that at early times, when $F_{\rm max}/F \gtrsim 1$, as long as $E_{\rm p}$ is above the XRT-band, then $\alpha_{\rm XRT} \sim |\alpha|$. 
In the case of $\gamma =1.5$ in the dissipative model, $E_{\rm p}$ rapidly crosses the entire XRT-band, which means that $\alpha_{\rm XRT} \sim |\beta|$ at late times. How quickly this transition occurs depends on the initial peak energy: if the peak energy is high, one probes the low-energy part for longer, and the transition occurs later. 

According to the \citet{Ronchini2021} analysis, $\alpha_{\rm XRT}$ seems to saturate around $\sim 2.3$ once $F_{\rm max}/F \gtrsim 100$. If $E_{\rm p} \ll 0.5~$keV at late times, this requires the existence of a persistent high-energy power-law with slope $\beta \sim -2.3$. This feature is non-trivial to maintain in a photospheric framework, since if dissipation halts deep below the photosphere, the high-energy photons lose their energy due to Compton scattering. Inelastic scatterings between neutrons and protons may provide such a signature, however, this requires a highly relativistic jet and may, therefore, not be efficient at late times in the current framework \citep{Beloborodov2010}. We note that the $\alpha-F$relation can be obtained if $\gamma$ is time-dependent in such a way that the decrease in $E_{\rm p}$ halts just as $\alpha_{\rm XRT}\sim 2.3$. However, unless one can find a physical motivation for why this would occur, this scenario seems unlikely. Therefore, we conclude that dissipation should continue all the way to the photosphere to accommodate the late-time spectrum (however, see discussion in Sect. \ref{sec:caution}).

If $\alpha_{\rm XRT}\sim |\alpha|$ at early times and $\alpha_{\rm XRT} \sim |\beta|$ at late times, then the dispersion observed in $\alpha_{\rm XRT}$ should reflect the dispersions found for $\alpha$ and $\beta$ in GRB catalogues. Specifically, the spread in $\alpha_{\rm XRT}$ at late times should increase, since the values obtained for $\beta$ in time-resolved analyses of the prompt emission ranges from $-4$ to $-2$ \citep{Poolakkil2021}. Thus, the perceived saturation around $\sim 2.3$ is likely due to the small number of GRBs in the current sample. If the behaviour persists as more GRBs are added to the sample, an alternative explanation for the power-law index at late times may be needed.


\subsection{Insights regarding the progenitor systems}\label{sec:progenitor}
To account for the observations using the model presented herein, a power-law decay of the kinetic power with an index $\sim 3-5$ is required. In this section, we briefly discuss what this implies for the dying central engine.

Many detailed numerical simulations of compact binary mergers have been published in the wake of GW170817. 
Although these numerical simulations cannot yet probe the long timescales discussed here (several $100~$s), they can still provide valuable insights. One such insight is that the mass accretion rate onto the black hole after collapse seems to decrease as $\sim t^{-2}$ \citep{Christie2019, MetzgerFernandez2021, Hayashi2023, Gottlieb2023arXiv}. Accounting for a conversion efficiency between the accretion rate and the jet kinetic power, as well as the uncertainty in $\epsilon_{\gamma}$, a decay of the kinetic power as ${\dot E}_{\rm K} \propto t^{-3}$ seems possible. This also motivates our choice of a power-law decay for the jet kinetic power in Eq. \eqref{eq:time_evolution}.

For collapsars, the picture is less clear. The mass accretion rate depends on the surrounding stellar density profile, $\rho(r) \propto r^{-a}$. Assuming free-fall accretion, the black hole mass accretion rate goes as $\sim t^{1-2a/3}$, with the stellar material at an initial radius $R$ reaching the central black hole of mass $M_{\rm BH}$ at a characteristic time $t = R^{3/2}/\sqrt{2G M_{\rm BH}}$ \citep[e.g.,][]{Gottlieb2022}. Assuming perfect conversion between accreted mass rate and jet kinetic power, a power-law decrease of ${\dot E}_{\rm K} \propto t^{-3}$ after $t=20~$s requires $a \sim 6$ at $R > 7.5 \times 10^9~$cm. 

A density decrease with $a \sim 6$ is much steeper than what is expected within a stellar envelope \citep[e.g.,][]{WoosleyHeger2006}. Indeed, it was recently found that the density profile after core collapse is quite shallow in the inner regions with $a \sim 1.5$ \citep{Halevi2023}. However, in the same work, the density was found to decrease much more rapidly close to the stellar edge, at $R \gtrsim 3 \times 10^9~$cm. Such a combined stellar density distribution would lead to a jet power that is initially constant, followed by a strong decay. Furthermore, it is sufficient that the diffuse material extends up until $\gtrsim 3 \times 10^{10}~$cm. This would generate an accretion rate $\propto t^{-3}$ over a timescale of $\sim 200~$s in the central engine frame, which is further stretched by a factor of $(1+z)$ in the observer frame. Thus, we conclude that the envisioned scenario predicts a diffuse density profile with $a \sim 6$ near the stellar edge in collapsars, which should extend to a few times $10^{10}~$cm. Since the ESD is a robust feature, these progenitor properties should be quasi-universal, which, therefore, constitutes a powerful test for the model.

The kinetic power is related to the Lorentz factor as ${\dot E}_{\rm K} = \Gamma {\dot M} c^2$, where ${\dot M}$ is the observed mass ejection rate in the jet. If the comoving density in the jet is constant, then ${\dot M} \propto \Gamma$ and we naturally obtain $\Gamma \propto t^{-1.5}$. In the second considered case, where $\Gamma \propto t^{-7/8}$, we require a comoving jet density that decreases with time. This could possibly be due to a cleared jet funnel resulting in less mixing at later times. 

\subsection{Photospheric blending}\label{sec:blending}
The ESD is often smooth in time, even if the earlier prompt emission has been highly irregular and chaotic. This may seem to contradict the scenario discussed in this paper: if the ESD probes the dying central engine, surely some variability is expected. However, a variable central engine will inevitably generate emission periods with higher optical depths, leading to larger photospheric radii. These dense shells are going to retrap emission from inner layers that may already be optically thin. The radiation from these different layers will blend and be emitted together at a later time. This effect ensures that the central engine variability is smoothed out in the observer frame.

Imagine a dense shell emitted at a time $t_1$. A second shell emitted at a later time $t_2 \equiv t_1 + \delta t$ will be affected by photospheric blending if the radiation released from the photosphere of the second shell reaches the first shell while it is still optically thick. This gives a condition on $\delta t$ as

\begin{equation}\label{eq:shadow}
	\delta t < \frac{R_{{\rm ph}, 1}}{\beta_1 c}(1-\beta_1) - \frac{R_{{\rm ph}, 2}}{\beta_2 c}(1-\beta_2) \approx \tau_{{\rm geo},1} - \tau_{{\rm geo},2}.
\end{equation}

\noindent Here, $R_{{\rm ph}, i}$, $\beta_i$, and $\tau_{{\rm geo},i}$  are, respectively, the photospheric radius, velocity, and geometrical timescale of the shell emitted at $t_i$, and the last approximation holds if $\Gamma_1,\Gamma_2 \gg 1$. If the central engine is variable at the end of its life, the ESD would deviate from a single power law in time. However, small-scale variations of timescales $\delta t$ would be suppressed in the light curve. 

Rapid variability is not observed during the EDS but X-ray flares, with typical timescales of $\delta t/t_{\rm obs} \sim 0.1$ to $1$, are quite common \citep{Burrows2005, Nousek2006}. Photospheric blending could be an interesting avenue to explore with regards to X-ray flares, since $\delta t \sim \tau_{\rm geo}$. In the dissipative case when $\gamma = 1.5$, one has $\tau_{\rm geo} \lesssim t_{\rm obs}$ after $\sim 400~$s. Therefore, at late times we have $\delta t \lesssim t_{\rm obs}$. Thus, photospheric blending may potentially act as a filter, suppressing small timescale variability while leaving large timescale variability unaffected. We leave a detailed investigation for a future time.

The effect of photospheric blending is general to photospheric models. Specifically, it should be present during the prompt emission as well. However, it is likely negligible during this phase as the geometrical timescale is expected to be small for standard GRB parameters. It becomes significant in cases when $\Gamma$ is small, e.g., when $t \gg t_{\rm b}$ in the current framework, or in cases when the kinetic power is exceptionally high.

\subsection{Assumptions in the previous works}\label{sec:caution}
In this section, we mention some underlying assumptions in the works of \citet{Ronchini2021} and \citet{Tak2023}, which may influence our results.

In \citet{Tak2023}, the peak energy is found to decrease with time as $E_{\rm p} \propto t^{-1}$ during the decay phases of a large fraction of the GRB pulses studied. This decline is in agreement with the theoretical expectation from HLE. However, in their study, only data from the Fermi Gamma-ray Burst Monitor (GBM) is used. 
The low-energy threshold of GBM is $8$~keV, with full effective area above $\sim 20$~keV \citep{Meegan2009}. 
Therefore, $E_{\rm p}$ remains clearly visible in the GBM only a short time after the peak of the pulse. Thus, the performed analysis can not probe very deep into the HLE-regime, with the peak energy in some of the studied pulses being tracked for $\lesssim 10~$s. Since the behaviour $E_{\rm p} \propto t^{-1}$ is expected only once the line-of-sight contribution has faded, a robust conclusion about the decay rate of the peak energy may require additional observations in lower-energy bands. This may be possible for  GRBs detected by the future space mission SVOM to be launched in 2024 \citep{Wei2016}. Its two gamma-ray instruments ECLAIRs and GRM offer a spectral coverage of the prompt emission from 4 keV to 5 MeV \citep{Bernardini2017}. 

In Fig. \ref{fig:alpha_F}, the evolution of the model spectra in the XRT-band during the ESD is shown in comparison with data points obtained by \citet{Ronchini2021}. However, such a comparison is not straightforward. The data points are generated by fitting a power-law function to the XRT spectral data during the ESD, and, thus, they include instrumental effects and background noise. The evolution of the photon index for the models on the other hand, is calculated using Eq. \eqref{eq:alpha}, following the definition of $\alpha_{\rm XRT}$ in \citet{Ronchini2021}. Eq. \eqref{eq:alpha} estimates the photon index by approximating the model spectra between 0.5~keV and 10~keV by a single power law (i.e., a straight line across the purple shaded region in Fig. \ref{fig:spectra}). Such a prescription neglects the shape of the model spectrum within the XRT-band. A more fair comparison against the data points would be achieved if one instead fitted a power-law function to mock data, where the mock data is generated by folding the model spectra through the XRT response matrix. It is plausible that the spectral index obtained using this method would differ from that obtained using Eq. \eqref{eq:alpha}, especially in regimes of low signal-to-noise. 

Lastly, the photon index obtained when fitting the XRT-data is highly sensitive on the modelling of the X-ray absorption below $\lesssim 1~$keV. Absorption of X-rays occurs in the Milky Way, in the host galaxy of the GRB, and possibly also in the intergalactic medium \citep{Wilms2000, Behar2011}. X-ray spectra from distant sources are commonly fitted with an absorbed power-law, where the absorption is modelled with a galactic plus an intrinsic hydrogen column density, $N_H$ \citep{Starling2013}. The absorbed power-law model often gives good fits to the data. However, the value of the photon index obtained is sometimes highly degenerate with the best-fit value for the hydrogen column density \citep[e.g.,][]{Valan2023}. This point is indeed raised and discussed in \citet{Ronchini2021}, who argue that their general results are robust against such a degeneracy. However, it highlights the importance of a correctly modeled absorption. 
Because of this, and the argumentation in the paragraph above, one should interpret the observational results presented in Fig. \ref{fig:alpha_F} with some caution.

\section{Conclusion}\label{sec:conclusion}
In this paper, we have investigated the end phase of the prompt emission in photospheric models of GRBs. Due to the small geometrical timescale expected in these models for typical GRB parameter values, the interpretation of the ESD as HLE emission is challenging. Instead, we interpreted the ESD as an emission signature from the dying central engine. 

We modelled the fading central engine by prescribing a power-law decay for the injected power, ${\dot E}_{\rm K}$, and the Lorentz factor, $\Gamma$, and constructed simple non-dissipative and dissipative frameworks to obtain the observed spectrum as a function of time (Fig. \ref{fig:spectra}). In the dissipative case, we found that the photospheric emission from the dying central engine can mimic the spectral evolution predicted from HLE, if the kinetic power and the Lorentz factor decrease as $t^{-3}$ and $t^{-7/8}$, respectively. If the Lorentz factor decreases more rapidly, as $t^{-1.5}$, the dissipative model can reproduce the $\alpha-F$ relation obtained by \citet{Ronchini2021} (Fig. \ref{fig:alpha_F}). This requires the existence of a persistent, high-energy power law to account for the spectral shape at late times, indicating that dissipation should take place near the photosphere. In both cases, we found that if the kinetic power decreases more quickly than $t^{-5}$, some fine-tuning of the other model parameters is necessary to be consistent with the observations. These results rely on a series of simplifying assumptions. A more detailed approach could include a time-varying efficiency and a physically motivated calculation of the spectral shape.

In Sect. \ref{sec:progenitor}, we used the deduced late-time behaviour of the central engine to gain insights about the progenitor systems. We argued that the jet kinetic power decreasing as $t^{-3}$ is in rough agreement with current state-of-the-art numerical simulations of compact binary mergers. If the comoving density in the jet remains constant, this in turn implies that $\Gamma \propto t^{-1.5}$. For collapsars, the evolution ${\dot E}_{\rm K} \propto t^{-3}$ after the initial prompt phase requires a diffuse density profile near the stellar edge, where the mass density $\rho \propto r^{-a}$ decreases rapidly with $a \sim 6$. This assumed free-fall accretion. The diffuse structure should extend from $R \lesssim 10^{10}~$cm to $R \gtrsim 3\times 10^{10}~$cm to account for the observed duration of the ESD.

Lastly, we discussed photospheric blending: the fact that outer regions of the jet with high optical depths may obscure emission from inner, optically thin layers, leading to a blending of the radiation from the different regions. Although a generic feature in photospheric models of GRBs, this effect is likely negligible during the prompt phase for standard GRB parameter values, since its typical duration scales with the geometrical timescale of the dense regions (Eq. \eqref{eq:shadow}). 
However, it should help suppress short timescale variability during the ESD in the current context, since the geometrical timescale can become significant at late times.

To conclude, spectral and temporal observations of the ESD can be reproduced by late-time photospheric emission from a dying central engine. To account for the observed light curve, we have found that the model predicts a decline of the kinetic power with a temporal index between $\sim -3$ to $-5$. A decline of the Lorentz factor with a temporal index $\sim -7/8$ reproduces a spectral behaviour similar to HLE, while a temporal index of $\sim -1.5$ reproduces the $\alpha-F$ relation. The behaviour should be quasi-universal to GRB central engines during their last stages and the model predictions can be tested against long-lasting numerical simulations in the future.\\[0.5cm]
We thank the anonomous referee for many good suggestions and the XRT-team for providing the data and light curves. F.A. is supported by the Swedish Research Council (Vetenskapsr\aa det; 2022-00347).
F.D. and R.M. acknowledge the Centre National d’Études Spatiales (CNES) for financial support in this research project.


\bibliographystyle{mnras}
\bibliography{References}
\end{document}